\documentclass[10pt,final, conference,letterpaper]{IEEEtran}

\usepackage{booktabs} 
\usepackage{amsmath}
\usepackage{graphicx}
\usepackage{algorithm,algpseudocode}
\usepackage{amsmath}
\usepackage{amsfonts}
\usepackage{amssymb}
\usepackage{bbding}
\usepackage{pifont}
\usepackage{wasysym}
\usepackage{amssymb}

\title{\fontsize{21.5}{25.8}\selectfont PUF-RLA: A PUF-based Reliable and Lightweight Authentication Protocol employing Binary String Shuffling}

\author
  {
	  \IEEEauthorblockN
    {
    Mahmood Azhar Qureshi\IEEEauthorrefmark{1} and
    Arslan Munir\IEEEauthorrefmark{2}
    }
    \IEEEauthorblockA
    {
    Department of Computer Science, Kansas State University \\
    Email:                                        %
    \IEEEauthorrefmark{1}mahmood102@ksu.edu, and %
    \IEEEauthorrefmark{2}amunir@ksu.edu
    \vspace{-3 mm}
    }
  }

\begin{document}

\maketitle
\thispagestyle{empty}
\pagestyle{empty}

\begin{abstract}
Physically unclonable functions (PUFs) can be employed for device identification, authentication, secret key storage, and other security tasks. However, PUFs are susceptible to modeling attacks if a number of PUFs' challenge-response pairs (CRPs) are exposed to the adversary. Furthermore, many of the embedded devices requiring authentication have stringent resource constraints and thus require a lightweight authentication mechanism.
We propose PUF-RLA, a PUF-based lightweight, highly reliable authentication scheme employing binary string shuffling. The proposed scheme enhances the reliability of PUF as well as alleviates the resource constraints 
by employing error correction in the server instead of the device without compromising the security. The proposed PUF-RLA is robust against brute force, replay, and modeling attacks. In PUF-RLA, we introduce an inexpensive yet secure stream authentication scheme inside the device which authenticates the server before the underlying PUF can be invoked. This prevents an adversary from brute forcing the device's PUF to acquire CRPs essentially locking out the device from unauthorized model generation. Additionally, we also introduce a lightweight CRP obfuscation mechanism involving XOR and shuffle operations. Results and security analysis verify that the PUF-RLA is secure against brute force, replay, and modeling attacks, and provides $\sim$99\% reliable authentication. In addition, PUF-RLA provides a reduction of 63\% and 74\% for look-up tables (LUTs) and register count, respectively, in FPGA compared to a recently proposed approach while providing additional authentication advantages.
\end{abstract}


\maketitle

\vspace{-4 mm}
\section{Introduction}
\vspace{-1 mm}
Ever since the introduction of physically unclonable functions (PUFs) \cite{Gassend:2002:SPR:586110.586132} more than one and a half decade ago, extensive research has been done in using these uncontrollable manufacturing variations for enhancing the device security. With the advent of advanced machine learning (ML)-based modeling techniques \cite{ruhrmair2010modeling}, strong PUFs (SPUFs), previously considered secure, now have their security in question. Given a number of challenge-response pairs (CRPs) of a 64x64 \textit{Arbiter PUF} \cite{majzoobi2009techniques}, an adversary can build a soft model for the device with a prediction accuracy of 99.9\% \cite{zhang2018dmos}. These model-building attacks can replicate the response behavior of the actual hardware and make the system vulnerable to attacks. Only eavesdropping the communication links back and forth, between the prover and the verifier, an adversary can still generate a highly accurate prediction model based on the past CRP exposure.  \par

Controlled PUFs (CPUFs) \cite{gassend2008controlled} are another class of SPUFs which enhance the security and resistance against ML-based modeling. These PUFs thwart model-building attacks by wrapping the PUF inside a control logic. 
One approach is to build the control logic in such a way as to limit the exposure of CRPs for the adversary \cite{yu2016lockdown,gao2018puf}. Another approach is to obfuscate the CRPs in such a way that even if the adversary can collect a number of CRPs, no effective model can be built since the original CRP relationship is only known to the device and the verifier \cite{zhang2018dmos}. \par

We propose a novel scalable authentication protocol which not only obfuscates the CRPs but also greatly limits the exposure of PUF responses thereby foiling any and all attempts made towards unauthorized model generation. 
In the previous works \cite{zhang2018dmos, yu2016lockdown, gao2018puf, rostami2012slender}, it was assumed that the server's database, containing the CRPs of the devices' PUF, is secure and an adversary has no capability to eavesdrop the communication link between the server and its database. For the previously proposed PUF-based security approaches \cite{zhang2018dmos, yu2016lockdown, gao2018puf, rostami2012slender}, if the data (CRPs or models) associated with the devices is compromised, then the security of the entire system would be breached. Also, with an increase in the number of devices to be authenticated by the server, storing the entire CRPs for all the devices in \textit{secure} server memory can be very costly. \par

Another major factor which affects PUF based authentication is the reliability of PUF. A PUF, for a given challenge, should ideally generate the same response under any environmental condition. This type of ideality is not possible in hardware as the response of a PUF to a particular challenge is dependent on the physical characteristics of the device. Under varying conditions (e.g., temperature, voltage) these physical characteristics differ resulting in generation of responses with bit flips associated with errors. Majority voting can help to reduce the errors but it does not guarantee high reliability under highly variable conditions. Contemporary approaches \cite{gassend2008controlled} and more recently proposed \cite{8353301} uses error correction codes (ECC) in the device as a fix for the PUF's reliability problem. These approaches do not consider the high hardware area overhead associated with computationally expensive error correction schemes in a low cost device. Moreover, ECC requires helper data to be communicated to the device by the server during an authentication round. This exposure of helper data provides another attack vector to the adversary who can use this information for modeling as well as side-channel analysis. We employ a different approach where the error correction is not present in the device, rather, the server is responsible for correcting the noisy responses of the device's PUF. By employing this, we not only guarantee a reliability of $\sim$99\% but also make the device extremely lightweight. Our main contributions are as follows:
%

%
\begin{itemize}
\item We employ a combination of binary string shuffling/deshuffling and XOR operations to obfuscate the CRPs associated with a device's PUF. No challenge or response in its original form is exposed on any communication link (i.e., between the device$\iff$server and server$\iff$database). Furthermore, the challenge and its issuance is strictly controlled by the device \textit{on-the-fly} within itself. Unlike some of the previous approaches which use cryptographic hash functions with high hardware overhead inside the device for response obfuscation, our scheme provides an extremely lightweight, low hardware cost dynamic obfuscation mechanism without compromising the security.
\item In PUF-RLA, the access to the underlying PUF in the device is strictly controlled. No challenge is issued and thus no response is generated unless a correct input stream is applied to the device, essentially locking the device from unauthorized access. This scheme is the first that locks out the device without even invoking the PUF, and thus completely inhibits any model-building as well as side-channel analysis attack on the underlying PUF.
\item PUF-RLA improves the PUF's response accuracy and therefore, the reliability, by employing error correction scheme in the server instead of the device. The device sends the noisy, obfuscated, shuffled responses to the server which, after de-obfuscating and deshuffling, corrects any underlying bit errors. This greatly reduces the hardware overhead of the devices in the system without compromising the reliability which makes this protocol easily deployable in low cost devices.
\end{itemize}

\vspace{-3 mm}
\section{Related Work}
\vspace{-1.5 mm}
Various schemes have been proposed in the past that implement a controlled strong PUF. Gassend \textit{et al.}\cite{gassend2008controlled} have proposed hashing of the input challenge and the response. However, this configuration requires hardware-expensive hashing as well as error correction logic in the device which makes it highly infeasible for low cost platforms. Also, the server in \cite{gassend2008controlled} needs to send the raw helper data to the device for stabilizing the noisy PUF responses. This exposes the PUF to attacks focusing on noise side-channel information \cite{becker2015pitfalls}. \par

Yu \textit{et al.} \cite{yu2016lockdown} have proposed an approach that upper bounds the available number of CRPs to an adversary. Only the trusted entity or the server can authorize the access of new CRPs. Furthermore, \cite{yu2016lockdown} has also introduced a nonce from the device-side to prevent reliability-based attacks \cite{becker2015pitfalls}. However, this approach supports only a limited number of authentication cycles (roughly 10,000) which makes it infeasible for IoT based deployments. Gao \textit{et al.} \cite{gao2018puf} have presented a finite state machine (FSM) locking mechanism at the output of the PUF circuit. A challenge is applied to the device and after evaluation, the responses from the PUF are fed to an FSM which traverses a given set of states till it reaches the final state. If a wrong input/challenge is applied to the PUF by an adversary, the response generated will prevent the FSM from reaching the final state thus not producing a valid response. The protocol presented in \cite{gao2018puf} seems to be sound but under strict ideal conditions (which is not a realistic assumption) where it is assumed that the device's PUF response will have a 0\% variation. This is because \cite{gao2018puf} hashes the output of the PUF response without error correction. Even a one bit error in the generated response during authentication will result in a wrong hash being computed and thus the protocol will fail under noisy conditions. Also, the inclusion of hash in the device drastically increases the hardware overhead. \par 

Rostami \textit{et al.} \cite{rostami2012slender} have introduced Slender PUF, which uses neither an error-correction logic nor any cryptographic protocol but it provides an open interface to the adversary. An adversary can acquire information about CRPs as long as the device's interface access is maintained. 
Chatterjee \textit{et al.} \cite{8353301} have recently published a work which uses PUF based authentication in an Internet of things (IoT) scenario and replaces the traditional certificate-based authentication. It is the first work in literature which considers the server's database to be breachable and secures it using keyed hash function. The server in \cite{8353301} only stores a single key in the non-volatile memory (NVM). However, during the authentication phase, the server sends raw challenges as well as the helper data associated to the PUF to the device. This exposure of helper data and challenges can result in side-channel attacks targeted on the device's PUF as the device's interface is open to random queries. Also, \cite{8353301} uses Bose-Chaudhuri-Hocquenghem (BCH) encoder/decoder based error correction logic inside the device to correct the noisy response which results in a high hardware area overhead. The protocols proposed in \cite{6881436}, \cite{Che2016APM} and  \cite{10.1007/978-3-662-48324-4_28} also have a huge area overhead which makes them unsuitable for low cost authentication purposes. \par 

In the proposed protocol, we close the device's interface from brute force queries by employing a stream authentication (SA) mechanism. We also improve the PUF's reliability by incorporating BCH-based error correction in the server. Unlike some previously proposed protocols, the presented work doesn't rely on cryptographic hashing for response obfuscation, rather it employs a combination of shuffling and XOR operations, which keeps the device lightweight and cost effective.
\begin{figure*}
\centering
\includegraphics[width=0.8\textwidth,keepaspectratio]{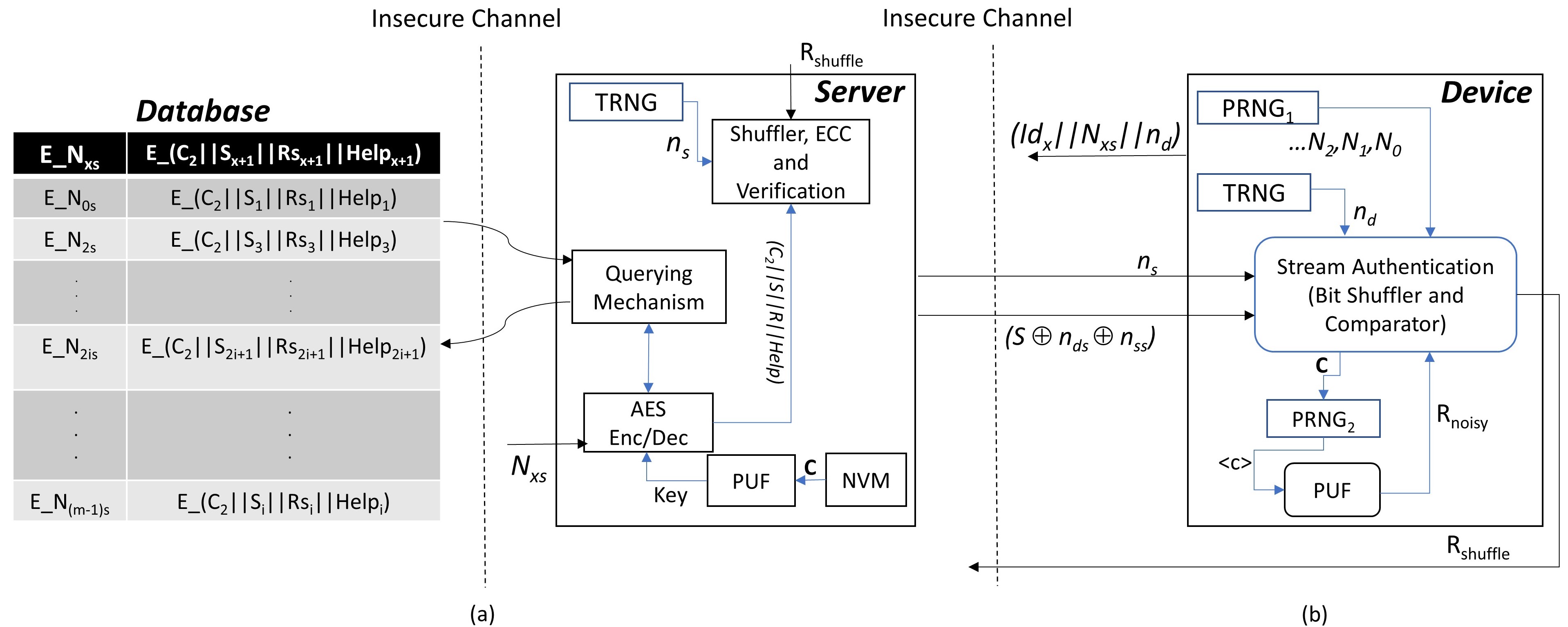}
\vspace{-3 mm}
\caption{Proposed scheme (a) Server-side along with the database, (b) Device-side}
\label{FigureProposedScheme}
\vspace{-5 mm}
\end{figure*}
\vspace{-3 mm}
\section{Proposed Scheme}
\vspace{-1 mm}
Figure~\ref{FigureProposedScheme} shows the high-level diagram of PUF-RLA for both the device- and the server-side. Both the channels, that is, between the server and the device and between the server and the database are insecure. Furthermore, the server also contains error correction block for correcting the noisy PUF responses. The server also incorporates a weak PUF which it uses to generate the encryption key for encrypting/decrypting the messages to and from the database. Instead of storing the key itself, the server stores a challenge to the PUF inside the NVM, which it uses to generate the $Key$. The device contains a \textbf{stream authentication (SA)} block comprising of a bit stream shuffler and a comparator. 

Similar to other authentication schemes \cite{zhang2018dmos, gassend2008controlled, yu2016lockdown, gao2018puf,  rostami2012slender,8353301}, this scheme has an \emph{enrollment phase} and an \emph{authentication phase}. The enrollment phase takes place in a secure environment during which the server assigns unique identifiers (IDs) to all the devices in the system. The assigned IDs are represented as $id_x$ where $x \in \mathbb{Z}^{+}$ denotes a device and $\mathbb{Z}^{+}$ denotes the set of positive integers (e.g., the ID of device $1$ is denoted as $id_1$). Device data (e.g., CRPs) is also collected during the enrollment phase in the secure environment.

\begin{algorithm}[!h]
    \caption{Enrollment Phase}
    \label{AlgoEnrollment}
    \begin{flushleft}
    \textbf{Input}: $SEED$: 128-bit initial seed value for $PRNG_1$ \\
    \textbf{Output}: Database containing encrypted concatenated values  \\
    \end{flushleft}
    \begin{algorithmic}[1]
    \Procedure{Enrollment}{} \
    \State $N_0,N_1,\ldots,N_i,\ldots,N_{m-1},N_m \gets PRNG_1(SEED)$
	\For{$i \gets 0$ to $m$} \
    \If{($i$ \textit{is even})}\
    \State $N_{is} \gets$ \textit{Shuffle}$(N_i)_{Counter_2}$ \
    \State $E\_N_{is} \gets Encrypt_{key}(N_{is})$ \
    \State \textit{Store $E\_N_{is}$ in} $i/2^{th}$ \textit{row and 1st column} \
    \Else
    \State $<c> \gets PRNG_2(N_i)$
    \State $Rs_i \gets$ ${Shuffle(PUF(<c>))_{Counter_2}}$\
    \State $Help_i \gets HelperDataGen()$\
    \State $S_i \gets$ \textit{Shuffle}$(N_i)_{Counter_1}$\
    \State $Encrypt_{key}(Counter_2||S_i||Rs_i||Help_i)$ \
    \State \textit{Store encrypted data in Ceil(i/2)th row and 2nd column} \
    \EndIf 
    \State  $Counter_1 \gets Counter_1 + CONST_1$ \
    \State $Counter_2 \gets Counter_2 + CONST_2$ \
    \EndFor
    \EndProcedure
    \end{algorithmic}
    \vspace{-1 mm}
\end{algorithm}
\vspace{-1 mm}
\subsection{Threat Model} \label{SectionThreatModel}
\vspace{-1 mm}
We consider a threat model where the authentication phase does not have to take place in a secure environment. The adversary, in our threat model, can eavesdrop, manipulate, or replay the traffic across all the communication links during authentication events. These communication links include the channel between the server and the devices as well as between the server and the database storing the device's data. By using these communication links, the adversary can attempt to model the device by utilizing the exchanged messages. The adversary can also perform \textit{man-in-the-middle} as well as spoofing attacks in order to gain unauthorized access. The device also has an open interface and the adversary can brute force query the device with any past or possibly adaptively chosen current messages/challenges.

\vspace{-2.5 mm}
\subsection{Enrollment Phase}
\vspace{-1 mm}

During the enrollment phase, a random number (RN) stream is generated by the $PRNG_1$ which uses a predetermined constant seed value as shown in Figure~\ref{FigureProposedScheme}(b). The total number of CRPs will be equal to the half of the total number of RNs generated. These random numbers are all 128 bits wide and expressed as:
\vspace{-1.5 mm}
%
%
\begin{equation}\label{eqRN}
RNs=N_0,N_1,\ldots,N_i,\ldots,N_{m-1},N_m,
\vspace{-1.5 mm}
\end{equation}
\vspace{-1 mm}
where $N_i$ is the $i^{th}$ random number. In our case, we set the maximum value of $i$, that is, $m$ to be $9999$ where $m$ must be odd (i.e., $m \equiv 1\ (\textrm{mod}\ 2)$) and less than equal to the period of the $PRNG_1$. We clarify that all the RNs in Eq.~\ref{eqRN} are unique (i.e., belong to a non-repetitive sequence) as $m \leq \text{period of } PRNG_1$. Every even-indexed RN is shuffled using a shuffling key represented by $Counter_1$ to get $N_{xs}$. The shuffled even-indexed RNs are encrypted and stored in the first column of the server's database. Every odd-indexed RN $(N_{x+1})$ is used as a challenge and passed as a seed for the $PRNG_2$ to derive the sub-challenges $(<c>)$ for the PUF in the device. The corresponding responses ($R_{x+1}$) are shuffled using $Counter_2$ to get shuffled responses ($Rs_{x+1}$). After the generation of $Rs_{x+1}$, the challenge (or odd-indexed RN ($N_{x+1}$)) is also shuffled using $Counter_1$ to get $S_{x+1}$. During the enrollment phase, \textit{helper data} $(Help_{x+1})$, associated with the device's PUF responses, is also generated. We then encrypt all the above generated variables to get $(enc(Counter_2||S_{x+1}||Rs_{x+1}||Help_{x+1}))$ which is stored in the second column of the server's database. In this way, for every value of encrypted $N_{xs}$ in the database, we will have a corresponding encrypted value containing the concatenation of the shuffling key $(Counter_2)$, the shuffled version of the challenge which is basically the odd-indexed RN $(S_{x+1})$, the shuffled version of the response of the PUF $(Rs_{x+1})$ and helper data $(Help_{x+1})$ corresponding to the particular response. This data is stored in the database as shown in Figure~\ref{FigureProposedScheme}(a).

%
Algorithm~\ref{AlgoEnrollment} shows the details of the enrollment in a more simplified and succinct manner. It can be seen that for every pair of even-odd indexed random numbers, the shuffling keys $(Counter_1, Counter_2)$ are different. This ensures that for every iteration, a different shuffling sequence is generated. It should also be noted that the shuffling key $Counter_1$ is not stored anywhere, rather it is used internally by the device during authentication. Therefore, the server itself cannot get the original challenges since the server does not store $Counter_1$. The server can only get the shuffled challenges from the database and send them to the device which uses $Counter_1$ to get the original challenge. This adds another layer of security as the original challenge is only known to the device itself. \par

After the completion of the enrollment phase, the device will have a predetermined seed value ($SEED$ in Algorithm 1), and $Counter_1$ and $Counter_2$ initialized to some initial value. After every successful authentication event, $Counter_1$ and $Counter_2$ will increment by a fixed number defined by $CONST_1$ and $CONST_2$ in Algorithm~\ref{AlgoEnrollment}.

\vspace{-2.5 mm}
\subsection{Authentication Phase}
\begin{figure}
\vspace{-2 mm}
\includegraphics[scale=0.49]{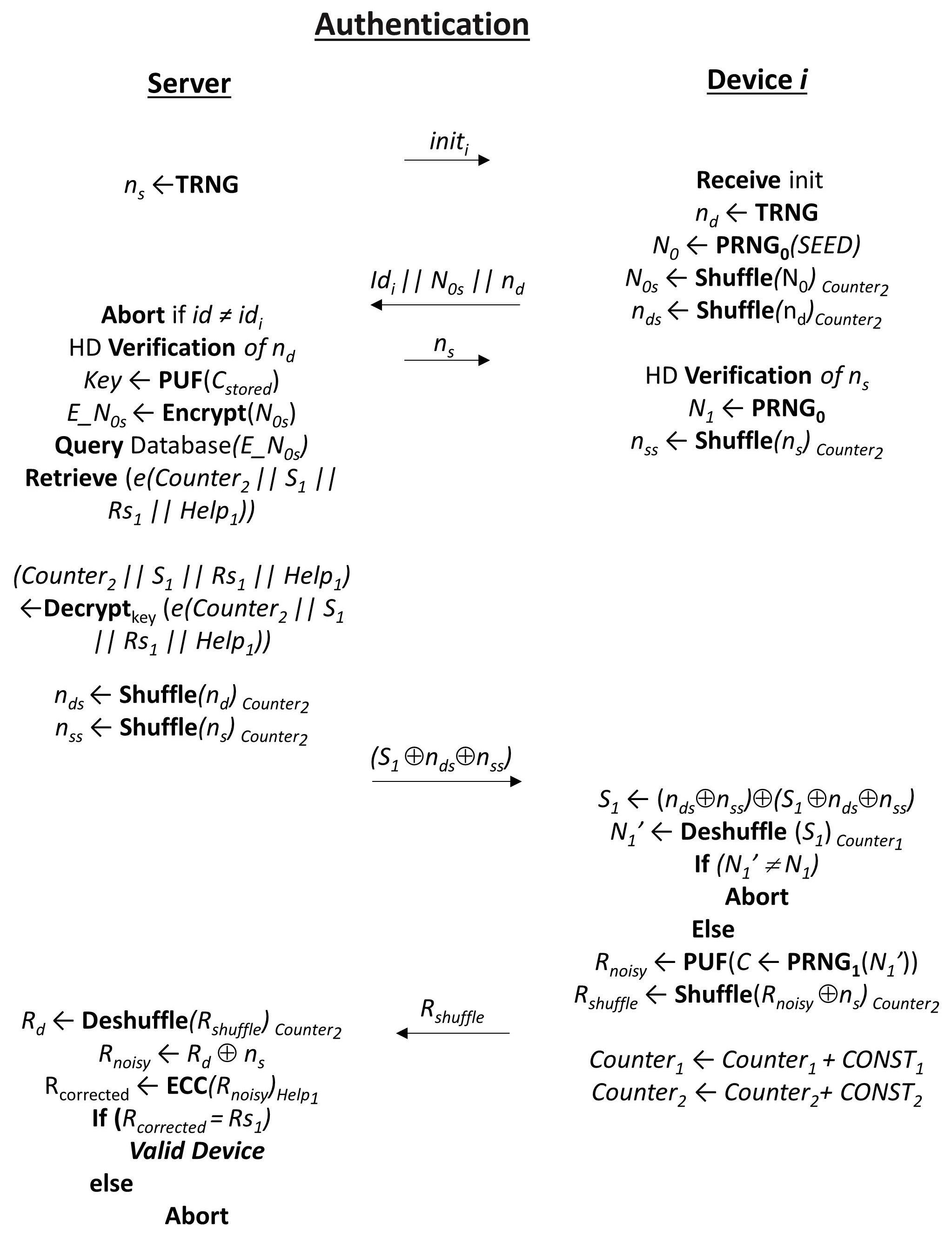}
\vspace{-3 mm}
\caption{Operational cycle of first authentication event}
\label{FigureAuthenticationProtocol}
\vspace{-9 mm}
\end{figure}

\vspace{-1.2 mm}
After successful enrollment of the device in the system, the next phase is the authentication phase. The authentication occurs in an insecure environment, where the communication channels can be monitored by untrusted parties. For better understanding of the protocol, we consider the first authentication cycle after the device has been deployed. Figure~\ref{FigureAuthenticationProtocol} shows the series of operations during the first valid authentication event between the server and the device. The server initiates the authentication session by sending an initialization message to the device. Upon sending the message, the server generates a random nonce $n_s$. The device upon getting the initialization command generates a random nonce $n_d$. The device also initializes the $PRNG_1$ with a predetermined $SEED$ and generates the first even-indexed RN $N_0$. This is the same as the first RN generated during the enrollment phase in Algorithm~\ref{AlgoEnrollment}. The device then uses the initial value of $Counter_2$ to shuffle $N_0$ and generates $N_{0s}$. The device then sends the concatenation of it's $ID$, the shuffled RN $N_{0s}$ and the nonce $n_d$ to the server. The device then shuffles the nonce $n_d$ using $Counter_2$ to get $n_{ds}$. The server upon receiving the message from the device verifies its ID and then performs a Hamming Distance (HD) check on the received $n_d$ from the device. HD is calculated between $n_d$ and an all zero string of length $L$, where $L$ is the bit-length of $n_d$. Similar HD calculation is performed between $n_d$ and all ones string .
This HD check is performed to ensure that $n_d$ has an acceptable distribution between ones and zeros. Our tests show that a $30/70$ ratio between zeros and ones and vice versa produces highly random shuffles with $50/50$ ratio being the ideal. \par

\begin{figure*}
\centering
\includegraphics[width=0.8\textwidth,keepaspectratio]{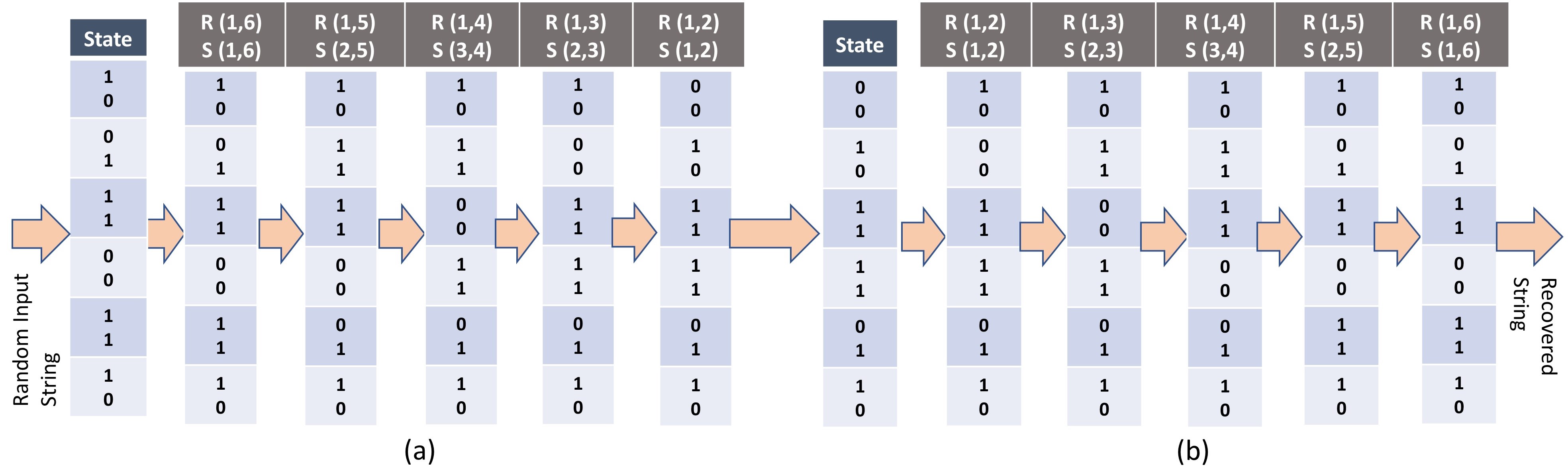}
\vspace{-3 mm}
\caption{(a) Shuffling of random input string, (b) Deshuffling to generate the original string}
\label{FigureShuffling}
\vspace{-4 mm}
\end{figure*}

After successful HD verification, the server sends it's own nonce $n_s$ to the device which performs the same HD check on $n_s$. The device then generates the next RN $N_1$ (odd-indexed) using $PRNG_1$. The device then shuffles $n_s$ using $Counter_2$ to get $n_{ss}$. Parallel to the computations in the device the server encrypts $N_{0s}$ using the encryption key and queries the database's first column to find the encrypted value. The server retrieves the encrypted value $(e(Counter_2||S_1||Rs_1||Help_1))$ corresponding to the entry $E\_N_{0s})$ from the database and decrypts it to get the individual components. $n_{ds}$ and $n_{ss}$ is generated in the server by shuffling $n_s$ and $n_d$ using the decrypted $Counter_2$ for this particular authentication. Note that this $Counter_2$ value in the server and the one in the device is the same. The server then sends the $XOR$ of the shuffled versions of the challenge (odd-indexed RN), $n_d$ and $n_s$ to the device. The device then retrieves the shuffled challenge $S_1$ by $XOR$-ing the received message with it's own $n_{ds}$ and $n_{ss}$. $S_1$ is then deshuffled using $Counter_2$ in the device to get $N_1'$. During a valid authentication event, that is, both the device and the server are trusted entities and without manipulation of the messages on the communication channel, $N_1'$ and $N_1$ (the odd-indexed RN generated in the device) should be the same. The device then uses $N_1'$ as a challenge to the PUF in the device to get a response $(R_{noisy})$, which depending on the environmental factors, can have bit errors. The device then shuffles the $XOR$ of the noisy response and $n_s$ and sends this to the server. The device then increments the two shuffling counters $Counter_1$ and $Counter_2$ by fixed constant values defined during the enrollment. The server upon receiving $R_{shuffle}$, performs deshuffling and $XOR$-ing to retrieve the noisy response $R_{noisy}$. The server then corrects the noisy response by using the error correction scheme and the helper data $Help_1$ that was retrieved from the database. The device is authenticated if the corrected response $R_{corrected}$ is the same as the response acquired from the database. 

After each authentication event, the device stores the most recently used odd-indexed $N_x$ for authentication (e.g., $N_1$ after the first authentication event) in a temporary memory so that in the next authentication event the device uses $N_{x+1}$ and $N_{x+2}$. For example, the device generates $N_2$ and $N_3$ as the second even- and odd-indexed RNs, respectively, using $PRNG_1$, and repeats the same cycle for authentication as depicted in Figure~\ref{FigureAuthenticationProtocol}. \par

This protocol also supports unlimited authentications which makes it an ideal candidate for IoT-based device authentications. After the authentication of the final pair of even-odd indexed RNs, the control logic resets the $PRNG_1$ to use the $SEED$ value again to start regenerating the same RNs. We will show in security analysis how, even after re-using the same RNs, the security of the protocol remains intact. \par
%

\vspace{-2.5 mm}
\subsection{Input Stream Authentication} \label{SectionStreamAuthentication}
\vspace{-1.6 mm}
The Stream Authentication (SA) block is the primary security primitive that separates the underlying PUF from the input. The SA block not only authenticates the stream issuing server but also prevents the device from responding in case an adversary tries to carry out a brute force or a replay attack. The SA block uses a bit shuffling/deshuffling scheme followed by a comparator. 
\vspace{-2mm}
\begin{algorithm}
    \caption{Bit Shuffling}
    \label{AlgorithmBitShuffuling}
    \begin{flushleft}
    \textbf{Input}: $n$: L-bit random number in binary format $({n_{L-1},n_{L-2},...,n_0})$  \\
    \textbf{Output}: $s$: L-bit shuffled version of the input  \\
    \end{flushleft}
    \begin{algorithmic}[1]
    \Procedure{Bit Shuffle}{} \
    \State $num_1,num_2,...,num_L \gets PRNG(shuffle_{key})$
	\For{$i \gets L$ to $1$} \
    \State $j \gets \textbf{Range}(num_{L-i+1},i)$
    \State \textbf{Swap} $(n_{j-1},n_{i-1})$
    \State $s_{L-1} \gets n_{j-1}$
    \EndFor
    \EndProcedure
    \end{algorithmic}
    \vspace{-1.5 mm}
    \end{algorithm}
\vspace{-1 mm}
\subsubsection{Bitstream Shuffler}
The proposed protocol employs a modified Fisher-Yates shuffling algorithm \cite{hazra2015file}. The algorithm produces an unbiased permutation, where every permutation is equally likely. Shuffling and deshuffling is performed in both the device and the server according to Figure 2. Algorithm~\ref{AlgorithmBitShuffuling} shows the working of the shuffling operation for any random number of bit length $L$.

\subsubsection*{\underline{Shuffling/Deshuffling Operation}} 
To perform shuffling, a $PRNG$ takes a shuffling key as input, which in our case is $Counter_1$ or $Counter_2$ in the device or $Counter_2$ in the server as shown in Figure~\ref{FigureAuthenticationProtocol}, and generates a number sequence $[num_1,num_2,\ldots,num_L]$ (line 2 of Algorithm 2). Here, $L$ is the bit length of the input number to be shuffled. The algorithm then iterates through all the numbers generated by the PRNG and swaps the $num_L$ indexed bit with the loop variable indexed bit (represented by $i$ in Algorithm 2). Proper range adjustment is also carried out to keep the indexing within bounds. By using the same algorithm in a reverse manner, that is, by using a shuffled number as input, the same shuffling key to generate the number sequence and iterating in a reverse manner, the original number can be recovered. \par


\subsubsection*{\underline{Illustrative Example of Shuffling and Deshuffling}} 
Figure~\ref{FigureShuffling} shows the process of shuffling and deshuffling on a small random test input by applying Algorithm~\ref{AlgorithmBitShuffuling}. Note that this is just a test case and the actual input is 128-bits wide. Also, for simplicity, we group together the input bits into two bit tuples and perform shuffling/deshuffling on the tuples instead of the individual bits in this illustration example (note that the shuffiling/deshuffling algorithm on actual 128-bits inputs is performed bitwise and not on tuples in PUF-RLA). An input bit string \textbf{\underline{100111001110}} corresponding to random number \textit{2510} is taken as an example. Assume that on providing a shuffling key to the PRNG, the number sequence comes out to be $[1,2,3,2,1,1]$. These numbers are adjusted to be in a decreasing range from \textit{R (1,6)}, with $1$ being the minimum number and $6$ being the maximum, to \textit{R (1,1)}, with $1$ being the only number in the range. \textit{R ($\cdot$,$\cdot$)} in Figure~\ref{FigureShuffling} represents the range of possible output values of the \textbf{Range($\cdot$,$\cdot$)} function (line 4) in Algorithm~\ref{AlgorithmBitShuffuling} for each corresponding iteration. For the first run of Algorithm~\ref{AlgorithmBitShuffuling}, the shuffling number is $1$, so we swap the $1^{st}$ tuple of bits with the last tuple (\textit{Swap(S) (1,6)}). In the next run, we decrease the range by one (i.e., from \textit{R (1,6)} to \textit{R (1,5)}). The shuffling number is $2$, so we swap the $2^{nd}$ and the $5^{th}$ tuple (\textit{S (2,5)}), and so on until we reach the last run (i.e., \textit{R (1,1)}) where no swapping is done. The final shuffled output comes out to be \textbf{\underline{001011110110}} corresponding to decimal \textit{758}. Note that for any other RN stream, the output will be different then the one presented above. Deshuffling is performed using the same number stream $[1,2,3,2,1,1]$ but in opposite manner (i.e., from \textit{R (1,1)} to \textit{R (1,6)}) to regenerate the original input as shown in Figure~\ref{FigureShuffling}(b). Note that the case of \textit{R(1,1)} is not shown in Figure~\ref{FigureShuffling} (a) and (b) as no shuffling/deshuffling is performed. In this way, if the server performs random shuffling of an $L$ bit message using a given key and sends the shuffled message to the device, the device can get the original message back by deshuffling using the same key. \par

It should be noted here that the security of the shuffling/deshuffling scheme relies on the fact that the number/message to be shuffled has at least 30/70 ratio between the number of ones and zeros. The RN sequence is carefully chosen during the enrollment phase (Algorithm~\ref{AlgoEnrollment}) by adjusting the $SEED$ value to ensure that the even- and odd-indexed RNs have somewhat equal distribution of zeros and ones. The shuffling keys ($Counter_1$ and $Counter_2$) are never sent over the communication channel and thus the adversary has no way of getting this information.

\subsubsection{Comparator}\label{SectionFSMbasedLocking}
The second sub-block in \textit{SA} is a comparator which checks whether the deshuffled input/challenge from the server equals the challenge generated by the device. The comparator also has control logic which, after determining the result of comparison ($TRUE$ or $FALSE$), enables or disables the underlying PUF circuit. If the comparison turns out to be $TRUE$, an $enable$ signal is sent to the $PRNG_2$ which uses the deshuffled validated challenge to generate sub-challenges $(<c>)$ for the PUF. If the input to the device has been manipulated, the comparator gives a $FALSE$ output and the $PRNG_2$ remains in disabled state, and thus the PUF circuit is not invoked and consequently does not generate any response. \par

Thus, by using a combination of the bit shuffler and the comparator, the SA block verifies the legitimacy of the server and prevents the device's PUF against targeted attacks focusing on brute force queries for the acquisition of responses for unauthorized modeling.

\begin{figure}
\centering
\includegraphics[width=0.35\textwidth,keepaspectratio]{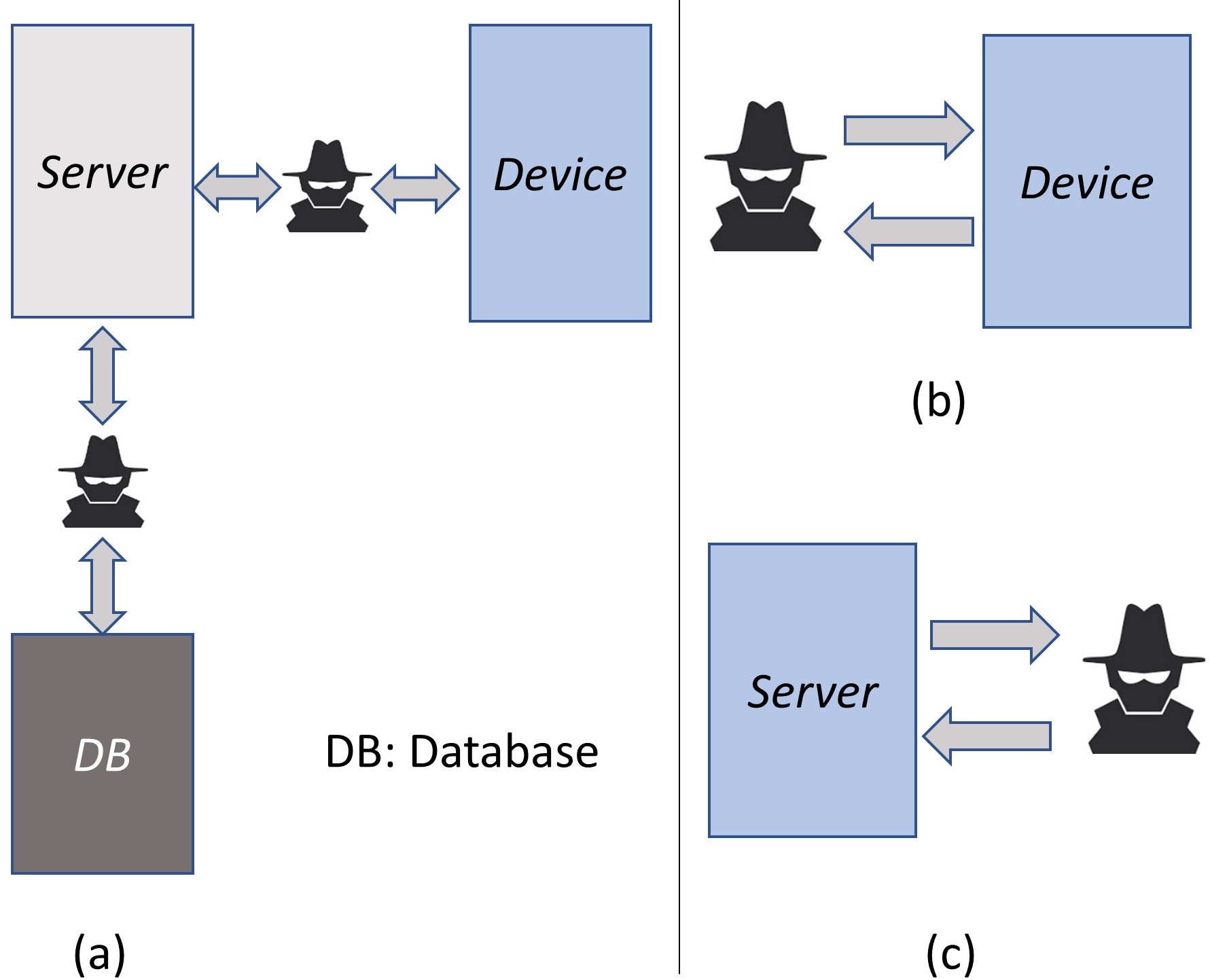}
\vspace{-3.2 mm}
\caption{Protocol-level attacks: (a) Man-in-the-middle attack, (b) Brute forcing the device, (c) Replay attack}
\label{FigProtocolAttacks}
\vspace{-5.2 mm}
\end{figure}
\vspace{-1 mm}
\section{Security Analysis}
\vspace{-1mm}
In this section, we analyze different attack scenarios by which the adversary can attempt to break the protocol. Like previously proposed schemes, we consider three main attacks which include \textit{man-in-the-middle} attack, brute force attack, and replay attacks as shown in Figure~\ref{FigProtocolAttacks}. As described in Section~\ref{SectionThreatModel}, the adversary in our case can not only monitor the communication channel between the server and the device but also between the server and it's database.

\vspace{-2mm}
\subsection{Man-in-the-middle Attack}
\vspace{-1 mm}
Man-in-the-middle (MITM) attack as shown in Figure~\ref{FigProtocolAttacks}(a) is a protocol-level attack in which the adversary secretly monitors the messages exchanged between the communicating parties. The adversary can delay, alter, or eavesdrop the messages over an insecure network. The communicating parties might think that their messages are being conveyed to the legitimate and intended parties, however, the messages are actually being manipulated by a third untrusted party. MITM is generally prevented by employing \textit{mutual authentication} where both parties mutually authenticate each other during message exchange. The protocols proposed in \cite{gao2018puf} and \cite{8353301} use hash function for mutual authentication however, the computation of hash is an expensive and time consuming operation. In PUF-RLA, we use shuffle and XOR operations for mutual authentication. Assuming that the adversary modifies $(S\oplus n_{ds}\oplus n_{ss})$ to $(S'\oplus n'_{ds}\oplus n'_{ss})_{MITM}$. The authentication of the server by the device, in this scenerio, will fail as:
\vspace{-1mm}
\begin{equation}
S_{device} \neq (S'\oplus n'_{ds}\oplus n'_{ss})_{MITM} \oplus  (S_{device}\oplus n_{ds}\oplus n_{ss})
\vspace{-0.5 mm}
\end{equation}
Since the adversary has no clue about $S$, $n_{ds}$, and $n_{ss}$, the only manipulation possible is the manipulation of combined XOR of the three components. The approximate probability of individually getting correct $(n_{ds}\oplus n_{ss})$ from $n_d$ and $n_s$ by brute forcing the shuffler is $\frac{1}{2^{127}} \times \frac{1}{2^{127}} \approx 3.45 \times 10^{-77}$ which is extremely low. Since the authentication takes place in real-time, the device will void the current authentication event if the correct $(S\oplus n_{ds}\oplus n_{ss})$ is not received from the server within some threshold time $\tau$\textit{ms}. The next authentication event is then initiated with next even- and odd-indexed RNs generated by $PRNG_1$ after some predetermined wait-time. After $\omega$ failed authentication events, the SA block locks out the device, which can only be unlocked explicitly by the server. This will prevent the adversary from trying out random input combinations. Similar rationale can be given for $R_{shuffle}$ where the unknown noisy response $R_{noisy}$ is XOR-ed with $n_s$ and then shuffled with a secret key. Only the server can extract a valid $R_{noisy}$ from $R_{shuffle}$ and disregard any manipulated $R_{shuffle}$. 

\vspace{-2 mm}
\subsection{Brute Force Attack}
\vspace{-1 mm}
Figure~\ref{FigProtocolAttacks}(b) shows the scenario where the adversary is brute forcing the device's interface to get the responses. The main goal of the brute force attack on PUF is to acquire data/CRPs of the PUF for modeling purposes. Since PUF-RLA restricts PUF's access at the input, the adversary cannot apply random challenges and get responses. The SA block will halt the operation, thereby not invoking the underlying PUF and thus no response will be generated by the device. This non-responsiveness of the device towards invalid inputs prevents the adversary from generating a soft model of the underlying PUF in the device. Also, the challenges issued by the server are shuffled which further restricts the adversaries' modeling capabilities. For the brute force attack, the attack point again is the manipulation of $(S\oplus n_{ds}\oplus n_{ss})$. Even though, both $n_s$ and $n_d$ are known to the adversary, generation of $(S\oplus n_{ds}\oplus n_{ss})$ for the SA block will prove to be a challenging and time consuming task since the shuffling key (i.e., $Counter_2$) is not known to the adversary. Also, the shuffled challenge $(S)$ can only be decoded by the device to recover the correct challenge for the PUF. Assuming uniform random distribution of 0's and 1's, the adversary will have to try out $2^{L}$ combinations, where $L$ is the bit length of the input. The adversary will again be locked out after $\omega$ failed authentication events. 

\vspace{-3 mm}
\subsection{Replay Attack} 
\vspace{-1 mm}
Replay attack is a type of MITM attack where the adversary uses previously exchanged valid authentication messages to masquerade as a legitimate entity. Generally, nonces or one-time session identifiers are used to protect against replay attacks. Figure~\ref{FigProtocolAttacks}(c) shows a scenario where the adversary masquerades as a legitimate device and performs a replay attack by using any previously sent messages over the communication links. Even if the adversary uses the previously sent messages for a particular valid authentication event, the server will still reject the authentication upon receiving $R_{shuffle}$ from the adversary. This is because $R_{shuffle}$ is generated by shuffling the XOR of the generated PUF response and the current nonce $n_s$ of the server. Since the adversary does not have the shuffling key $Counter_2$, and is using $R_{shuffle}$ generated from some previous $n_s$, the adversary cannot send a legitimate $R_{shuffle}$ for the current $n_s$ in the server. Thus, the server will treat any previously valid message as invalid for the current authentication cycle.

\vspace{-1 mm}
\section{Implementation and Results}
\vspace{-1 mm}
We use Xilinx $Zynq-7000$ ZC706 evaluation board, as the device, for the implementation of the proposed protocol. The programmable logic (PL) part of the Zynq board implements a finite state machine (FSM) which goes through all the authentication stages of the device presented in Figure~\ref{FigureAuthenticationProtocol}. The FSM calls all the major routines including SA, the PUF circuit, and the PRNGs. The processing system (PS) part of the Zynq board is responsible for the communication with a desktop/server running MATLAB R2018b via universal asynchronous receiver-transmitter (UART). Since the device is more resource constrained, we measure the hardware overhead of the device side of PUF-RLA with the device side of previous approaches.
\par 

\vspace{-2 mm}
\subsection{PUF}
\vspace{-1 mm}
The PUF used in this protocol is a variant of Arbiter PUF (APUF) known as Double Arbiter PUF (DAPUF). Implementation of PUFs, particularly arbiter-based PUFs, is challenging because of the strict symmetric routing constraints. Designers can do their best to do symmetric routing in FPGA fabric by imposing place and route constraints of various components. However, even after imposing these routing constraints, complete symmetry is still not achievable. Even a slight non-symmetry can cause a delay bias in the arbiter output which results in a loss of unpredictability of the PUF circuit. Authors in \cite{5711471} introduced the concept of implementing Arbiter PUFs using programmable delay lines (PDLs). We use the similar concept given in \cite{5711471} to implement a PDL-based 3-1 DAPUF using 6-input look-up tables (LUTs) in the PL part of Zynq-7000 SoC. PUFs are normally evaluated using three performance metrics: uniqueness, reliability, and randomness.

\begin{table}[ht]
\vspace{-4 mm}
\centering
  \caption{3-1 DAPUF Evaluation Metrics}
  \label{TablePUFEvalMetrics}
  \vspace{-2 mm}
  \begin{tabular}{ccccl}
    \toprule
    Property & Ideal & 3-1 APUF & 3-1 DAPUF\\
    \midrule
    Uniqueness (\%) & 50 & 6.34 & 51.7\\
    Reliability (\%) & 100 & 97.8 & 87.5\\
    Randomness (\%) & 50 & 54.33 & 53.2\\
    \bottomrule
  \end{tabular}
  \vspace{-2 mm}
\end{table}

Table~\ref{TablePUFEvalMetrics} shows a comparison of the various evaluation metrics of a 3-1 DAPUF and a basic 3-1 XOR APUF. The table is taken by averaging out the values generated after the analysis performed by \cite{6933107}.

\subsubsection{Uniqueness}
Uniqueness tells the amount of variation in PUF responses among different chips/instances. A $50$\% uniqueness is ideal which means that the hamming distance between responses of different PUF instances for the same challenge(s) differs by 50\% of total response bits. The uniqueness $\mathcal{U}$ for PUF-RLA is defined as:

\vspace{-5 mm}
\begin{equation}
\mathcal{U} = \dfrac{2}{nvr(r-1)} \sum_{p=1}^{r-1} \sum_{q=p+1}^{r} \sum_{i=0}^{v} HD\left(R_{p_{2i+1}},R_{q_{2i+1}}\right) \times 100\%, 
\end{equation}
where $v = (m/2 - 1)$ and $m$ is the last RN used in CRP evaluations as indicated in Eq.~(\ref{eqRN}), $r$ and $n$ are the total number of PUF instances and challenge/response bits, respectively. $R_{p_{2i+1}}$ and $R_{q_{2i+1}}$ denotes the PUF responses corresponding to the odd-indexed RN (challenge) for PUF instances $p$ and $q$, respectively. As shown in Table~\ref{TablePUFEvalMetrics}, the uniqueness of a basic 3-1 XOR APUF is very low which makes it unsuitable for authentication protocols. In comparison, the 3-1 DAPUF has a uniqueness much closer to the ideal value. \par

\subsubsection{Reliability}
For reliability testing, the PUF circuit is evaluated under varying conditions (e.g., voltage and temperature) and HD is calculated between the ideal and the obtained responses. Ideally, the same PUF instance should output the same response given a particular challenge. However, this is usually not the case and the generated responses have bit flips. The reliability $\mathcal{R}$ of PUF in PUF-RLA is given by:
\vspace{-2 mm}
\begin{equation}
\mathcal{R} = 100\% - \dfrac{1}{vnl} \sum_{i=0}^{v}\sum_{t=1}^{l}  HD\left(R_{2i+1},R_{2i+1,t}'\right) \times 100\%, 
\end{equation}
where $v = (m/2 - 1)$ and $m$ is the last RN used in CRP evaluations as indicated in Eq.~(\ref{eqRN}), $n$ is the number of bits in PUF response, $t$ represents the total number of distinct conditions (determined by temperature and voltage combination) for which the PUF response is evaluated for a given challenge, $R_{2i+1}$ denotes the PUF response corresponding to the odd-indexed RN (challenge), and $R_{2i+1,t}'$ is the $t^{th}$ sample of $R_{2i+1}$ obtained at a given condition (i.e., voltage and temperature value). Thus, reliability measure subsumes the average number of erroneous response bits for all the CRPs evaluated at varying temperature and voltage conditions. \par

In case of 3-1 DAPUF, the reliability is, on average, 13\% less then the ideal value and can go further down by, as high as, ~15\% \cite{6933107}. This is a very high error rate and can significantly impact the authentication reliability. Without error correction, the server and the device will have to restart the authentication round which can significantly increase the execution time of the protocol. Implementing error correction in the device, as was the case for all previous approaches employing error correction, significantly increases the area overhead. Furthermore, since error correction requires a particular $b$ number of \emph{syndrome/helper} bits, to be communicated to the device, an adversary can get $b$ bits about the PUF delay circuit. Since PUF-RLA provides the flexibility to implement error correction at the server end, in addition to not exposing the raw helper bits on any communication link, a good error correction scheme can be implemented in software on the server without increasing the device's area overhead. As a test case, we implement BCH encoder/decoder in software using MATLAB R2018b using the \textit{bchenc/bchdec} function. The implemented BCH decoder has the capability to correct up-to 20\% error rate. This accounts to a practical reliability of $\sim$99\% in addition to a $\sim$60\% decrease in the area overhead of the device. \par

\subsubsection{Randomness}
The final evaluation metric of a PUF is it's randomness. Randomness refers to the ratio of 0's and 1's in the PUF responses. Ideally, a PUF response should comprise of equal number of 0's and 1's. As shown in Table~\ref{TablePUFEvalMetrics}, the 3-1 DAPUF has, on average, a randomness of 53\% which is close to the ideal value. 
\par

\vspace{-2 mm}
\subsection{TRNG}
\vspace{-1 mm}
The 128-bit nonce $n_d$, generated by the device in PUF-RLA is through a true random number generator (TRNG). We use the approach presented in \cite{Majzoobi:2011:FTR:2044928.2044931} for generation of nonces in the PL of Zynq7000. The operation of this TRNG is governed by enforcing a metastable state on the flip-flop through a closed loop feedback control. Other than providing good randomness properties, the hardware footprint of this TRNG is very low making it an ideal candidate for generation of nonces in low cost authentication schemes. Interested readers can refer to \cite{Majzoobi:2011:FTR:2044928.2044931} for the details regarding this TRNG.

\vspace{-2 mm}
\subsection{Shuffler/Deshuffler}
\vspace{-1 mm}
The shuffling/deshuffling operation requires a 128-bit distributed RAM and a 128-bit temporary register which stores the intermediate results after every shuffle operation. A PRNG is used to generate a number sequence which uses a shuffling key as the seed value for the PRNG. The shuffler has a very low hardware footprint (roughly 100 LUTs and 200 flip-flops (FFs)) compared to cryptographic hashing schemes ($>$1000 LUTs and FFs \cite{rostami2012slender}). \par
\vspace{-2 mm}

\begin{table}
\centering
  \caption{Comparison against Previous Protocols}
  \label{TableSecurityComp}
  \vspace{-3 mm}
 \addtolength{\tabcolsep}{-1.2pt}
  \begin{tabular}{cccccccl}
    \toprule
    Property &\cite{gassend2008controlled} & \cite{rostami2012slender} & \cite{yu2016lockdown} & \cite{gao2018puf} & \cite{8353301} & PUF-RLA \\
    \midrule
    Scalable & \checkmark & \checkmark& \checkmark& \checkmark & \checkmark & \checkmark\\
    TRNG & \ding{55} & \checkmark & \ding{55} & \checkmark & \ding{55} & \checkmark\\
    Mutual Authentication & \ding{55} & \ding{55}& \checkmark& \checkmark & \checkmark & \checkmark\\
    Crypto Algo in the Device & \checkmark & \ding{55}& \ding{55} & \checkmark & \checkmark & \ding{55}\\
    Error Correction in Device & \checkmark & \ding{55}& \ding{55} & \ding{55} & \checkmark & \ding{55}\\
    Helper Data Exposed & \checkmark & \ding{55} & \ding{55} & \ding{55} & \checkmark & \ding{55}\\
    Authentication Rounds & $\infty$ & $\infty$ & \textit{d} & $\infty$ & $\infty$ & $\infty$\\
    Open Server-Database Link & \ding{55} & \ding{55} & \ding{55} & \ding{55} & \checkmark & \checkmark\\
    Open Device Interface & \checkmark & \checkmark & \ding{55} & \ding{55} & \checkmark & \ding{55}\\
    \bottomrule
  \end{tabular}
 \vspace{-4 mm}
\end{table}
%

\begin{table}[ht]
\centering
  \caption{Hardware Overhead Comparison}
  \label{TableOverhead}
  \vspace{-3 mm}
\addtolength{\tabcolsep}{3pt}
  \begin{tabular}{cccl}
    \toprule
    Protocols & \ LUT count & \ FF count \\
    \midrule
    \cite{gassend2008controlled} & not reported & not reported \\
    \cite{yu2016lockdown} & not reported & not reported \\
    \cite{gao2018puf} & 960 & 1500 \\
    \cite{rostami2012slender} & 652 & 537 \\
    \cite{8353301} & 1591 & 1933 \\
    \cite{6881436} & 9207 & 2921 \\
    \cite{Che2016APM} & 6034 & 1724 \\
    \cite{10.1007/978-3-662-48324-4_28} & 3543 & 1275\\
    PUF-RLA & 590 & 510  \\
    \bottomrule
  \end{tabular}
  \vspace{-5 mm}
\end{table}
\subsection{Comparison With Other PUF-Based Protocols}
\vspace{-1.5 mm}
Tables~\ref{TableSecurityComp} and ~\ref{TableOverhead} show the comparison of various properties and hardware overhead, respectively, between PUF-RLA and the past approaches. It can be seen that the PUF-RLA not only retains the cumulative advantages of all the previous approaches while circumventing their limitations, it also reduces the hardware overhead to a great extent without compromising the security and reliability. The LUT count for PUF-RLA is divided among all the modules as 384 (3-1 DAPUF + TRNG) + 10 (PRNGs) + 120 (Shuffler) + 76 (main control) = 590. The FF count is divided as 20 (PUF + TRNG) + 128 (PRNG) + 204 (Shuffler) + 158 (main control) = 510. \cite{rostami2012slender} has somewhat similar LUT and FF count but it lacks the very important feature of mutual authentication. In addition, PUF-RLA also opens the link between between the server and the database which \cite{rostami2012slender} considers secure and circumvents the MITM as well as other attacks possible on this link. \cite{8353301} provides advantages similar to PUF-RLA when it comes to authentication but PUF-RLA has $\sim$63\% and $\sim$74\% reduction in LUT and FF count, respectively. Also, PUF-RLA closes the open device interface by enforcing the SA block on the PUF input which \cite{8353301} does not. \cite{6881436,Che2016APM,10.1007/978-3-662-48324-4_28} have a huge area overhead, as shown in Table~\ref{TableOverhead}, which makes them unsuitable for low cost platforms.

\vspace{-2 mm}
\section{Conclusions}
\vspace{-1 mm}
In this paper, we have proposed PUF-RLA, a controlled PUF-based, authentication protocol, which (i) closes the open interface between the input and the PUF by implementing a strong control logic that denies the PUF's access to the adversaries, (ii) makes the scheme highly reliable by incorporating error correction in the server thereby not revealing any helper data on insecure channels, and (iii) reduces the hardware overhead drastically by incorporating a lightweight CRP obfuscation mechanism employing bit shuffling and XOR operations. The security analysis of PUF-RLA verifies that the PUF-RLA is secure against brute force, replay, and modeling attacks. Results reveal that PUF-RLA provides ~99\% reliable authentication. Additionally, PUF-RLA is lightweight providing a reduction of 63\% and 74\% for look-up tables (LUTs) and register count, respectively, in FPGA as compared to a recently proposed approach while furnishing additional authentication advantages. Our future goal is to extend PUF-RLA to incorporate a secure key establishment protocol. This would make PUF-RLA deployable in IoT based systems.
%

\newpage
\bibliographystyle{unsrt}
\bibliography{PUF_RLA_bibv1}

\begin{thebibliography}{10}

\bibitem{Gassend:2002:SPR:586110.586132}
Blaise Gassend, Dwaine Clarke, Marten van Dijk, and Srinivas Devadas.
\newblock Silicon physical random functions.
\newblock In {\em Proc. of the 9th ACM Conference on Computer and
  Communications Security (CCS)}, pages 148--160, New York, NY, USA, 2002.

\bibitem{ruhrmair2010modeling}
Ulrich R{\"u}hrmair, Frank Sehnke, Jan S{\"o}lter, Gideon Dror, Srinivas
  Devadas, and J{\"u}rgen Schmidhuber.
\newblock Modeling attacks on physical unclonable functions.
\newblock In {\em Proc. of the 17th ACM Conference on Computer and
  Communications Security (CCS)}, pages 237--249, Chicago, IL, USA, 2010.

\bibitem{majzoobi2009techniques}
Mehrdad Majzoobi, Farinaz Koushanfar, and Miodrag Potkonjak.
\newblock Techniques for design and implementation of secure reconfigurable
  {PUFs}.
\newblock {\em ACM Trans. on Reconfigurable Technology and Systems (TRETS)},
  2(1):5, 2009.

\bibitem{zhang2018dmos}
Jiliang Zhang, Lu~Wan, Qiang Wu, and Gang Qu.
\newblock {DMOS-PUF}: Dynamic multi-key-selection obfuscation for strong {PUFs}
  against machine learning attacks.
\newblock {\em arXiv preprint arXiv:1806.02011}, 2018.

\bibitem{gassend2008controlled}
Blaise Gassend, Marten~Van Dijk, Dwaine Clarke, Emina Torlak, Srinivas Devadas,
  and Pim Tuyls.
\newblock Controlled physical random functions and applications.
\newblock {\em ACM Trans. on Information and System Security (TISSEC)},
  10(4):3, 2008.

\bibitem{yu2016lockdown}
Meng-Day Yu, Matthias Hiller, Jeroen Delvaux, Richard Sowell, Srinivas Devadas,
  and Ingrid Verbauwhede.
\newblock A lockdown technique to prevent machine learning on {PUFs} for
  lightweight authentication.
\newblock {\em IEEE Trans. on Multi-Scale Computing Systems (TMSCS)},
  2(3):146--159, 2016.

\bibitem{gao2018puf}
Yansong Gao, Hua Ma, Said~F Al-Sarawi, Derek Abbott, and Damith~C Ranasinghe.
\newblock {PUF-FSM}: A controlled strong {PUF}.
\newblock {\em IEEE Trans. on Computer-Aided Design of Integrated Circuits and
  Systems (TCAD)}, 37(5):1104--1108, 2018.

\bibitem{rostami2012slender}
M~Rostami, M~Majzoobi, Farinaz Koushanfar, Dan~S Wallach, and Srinivas Devadas.
\newblock Slender {PUF} protocol: A lightweight, robust, and secure
  authentication by substring matching.
\newblock In {\em IEEE Symposium on Security and Privacy Workshops}, pages
  33--44, San Francisco, CA, USA, 2012.

\bibitem{8353301}
U.~{Chatterjee}, V.~{Govindan}, R.~{Sadhukhan}, D.~{Mukhopadhyay}, R.~S.
  {Chakraborty}, D.~{Mahata}, and M.~M. {Prabhu}.
\newblock Building {PUF} based authentication and key exchange protocol for
  {IoT} without explicit {CRPs} in verifier database.
\newblock {\em IEEE Transactions on Dependable and Secure Computing}, pages
  1--1, 2018.

\bibitem{becker2015pitfalls}
Georg~T Becker.
\newblock On the pitfalls of using arbiter-{PUFs} as building blocks.
\newblock {\em IEEE Trans. on Computer-Aided Design of Integrated Circuits and
  Systems (TCAD)}, 34(8):1295--1307, 2015.

\bibitem{6881436}
J.~{Kong}, F.~{Koushanfar}, P.~K. {Pendyala}, A.~{Sadeghi}, and C.~{Wachsmann}.
\newblock {PUFatt}: Embedded platform attestation based on novel
  processor-based {PUFs}.
\newblock In {\em 51st ACM/EDAC/IEEE Design Automation Conference (DAC)}, pages
  1--6, San Francisco, CA, USA, June 2014.

\bibitem{Che2016APM}
Wenjie Che, Mitchell Martin, Goutham Pocklassery, Venkata~K. Kajuluri, Fareena
  Saqib, and James~F. Plusquellic.
\newblock A privacy-preserving, mutual {PUF}-based authentication protocol.
\newblock {\em Cryptography}, 1:3, 2016.

\bibitem{10.1007/978-3-662-48324-4_28}
Aydin Aysu, Ege Gulcan, Daisuke Moriyama, Patrick Schaumont, and Moti Yung.
\newblock End-to-end design of a {PUF}-based privacy preserving authentication
  protocol.
\newblock In Tim G{\"u}neysu and Helena Handschuh, editors, {\em Cryptographic
  Hardware and Embedded Systems (CHES)}, pages 556--576. Springer Berlin
  Heidelberg, 2015.

\bibitem{hazra2015file}
Tapan~Kumar Hazra, Rumela Ghosh, Sayam Kumar, Sagnik Dutta, and Ajoy~Kumar
  Chakraborty.
\newblock File encryption using {Fisher-Yates} shuffle.
\newblock In {\em International Conference and Workshop on Computing and
  Communication (IEMCON)}, pages 1--7, Vancouver, British Columbia, Canada,
  2015.

\bibitem{5711471}
M.~{Majzoobi}, F.~{Koushanfar}, and S.~{Devadas}.
\newblock {FPGA PUF} using programmable delay lines.
\newblock In {\em IEEE International Workshop on Information Forensics and
  Security}, pages 1--6, Seattle, WA, USA, Dec 2010.

\bibitem{6933107}
T.~{Machida}, D.~{Yamamoto}, M.~{Iwamoto}, and K.~{Sakiyama}.
\newblock A new mode of operation for arbiter {PUF} to improve uniqueness on
  {FPGA}.
\newblock In {\em Federated Conference on Computer Science and Information
  Systems}, pages 871--878, Warsaw, Poland, Sep. 2014.

\bibitem{Majzoobi:2011:FTR:2044928.2044931}
Mehrdad Majzoobi, Farinaz Koushanfar, and Srinivas Devadas.
\newblock {FPGA}-based true random number generation using circuit
  metastability with adaptive feedback control.
\newblock In {\em Proceedings of the 13th International Conference on
  Cryptographic Hardware and Embedded Systems (CHES)}, pages 17--32, Berlin,
  Heidelberg, 2011. Springer-Verlag.

\end{thebibliography}

\end{document}